\documentclass{article}%
\usepackage{amsmath}
\usepackage{graphicx}
\usepackage{amsfonts}
\usepackage{amssymb}
\usepackage{geometry}%
\providecommand{\U}[1]{\protect\rule{.1in}{.1in}}

\DeclareFontFamily{OT1}{pzc}{}
\DeclareFontShape{OT1}{pzc}{m}{it}{<-> s * [1.10] pzcmi7t}{}
\DeclareMathAlphabet{\mathpzc}{OT1}{pzc}{m}{it}
\providecommand{\U}[1]{\protect\rule{.1in}{.1in}}
\providecommand{\U}[1]{\protect\rule{.1in}{.1in}}
\providecommand{\U}[1]{\protect\rule{.1in}{.1in}}
\providecommand{\U}[1]{\protect\rule{.1in}{.1in}}
\providecommand{\U}[1]{\protect\rule{.1in}{.1in}}

\newcommand{\bee}{\begin{equation}}
\newcommand{\eend}{\end{equation}}

\newcommand{\bea}{\begin{eqnarray}}
\newcommand{\eea}{\end{eqnarray}}

\begin{document}

\title{Finite field-energy of a point charge in QED}
\author{Caio V. Costa$^{1}$\thanks{Electronic address: caiocostalopes@usp.br}, Dmitry
M. Gitman$^{1}$\thanks{Electronic address: gitman@dfn.if.usp.br} and Anatoly
E. Shabad$^{2}$\thanks{Electronic address: ashabad@Ipi.ru}\\$^{1}$\textsl{Instituto de F\'{\i}sica, Universidade de S\~{a}o Paulo, }\\\textsl{Caixa Postal 66318, CEP 05508-090, S\~{a}o Paulo, S. P., Brazil} \\$^{2}$\textsl{P. N. Lebedev Physics Institute, Leninsky Prospekt 53, Moscow
117924, Russia} }
\maketitle

\begin{abstract}
We consider a simple nonlinear (quartic in the fields) gauge-invariant
modification of classical electrodynamics, which possesses a regularizing
ability sufficient to make the field energy of a point charge finite. The
model is exactly solved in the class of static central-symmetric electric
fields. Collation with quantum electrodynamics (QED) results in the total
field energy about twice the electron mass. The proof of the finiteness of the
field energy is extended to include any polynomial selfinteraction, thereby
the one that stems from the truncated expansion of the Euler-Heisenberg local
Lagrangian in QED in powers of the field strenth.

\end{abstract}

\bigskip

\section{Introduction: the quartic model}

\bigskip We define the Lagrangian of a minimally nonlinear electrodynamics as%
\begin{equation}
L(x)=-\mathfrak{F}\left(  x\right)  +\frac{\gamma}{2}\left(  \mathfrak{F(}%
x)\right)  ^{2}, \label{L}%
\end{equation}
where $\mathfrak{F}\left(  x\right)  =\frac{1}{4}F^{\mu\nu}F_{\mu\nu}=\frac
{1}{2}\left(  B^{2}-E^{2}\right)  $ is the first electromagnetic field
invariant that makes (with the reversed sign) the Lagrangian density of the
standard linear Maxwell electrodynamics, while the second term in (\ref{L}) is
the quartic in the field-strength addition to it\footnote{Greek indices span
the 4-dimensional Minkowski space-time taking the values 0,1,2,3, while the
Roman indices are 1,2,3. The metric tensor is $\eta_{\mu\nu}=\mathrm{diag}%
(-1,+1,+1,+1),$ and bold symbols are reserved for three-dimensional Euclidean
vectors. The Heaviside-Lorentz system of units is used throughout.}. The
field-strength tensor is related to the four-vector potential $A^{\mu}\left(
x\right)  $ as $F^{\mu\nu}=\partial^{\mu}A^{\nu}(x)-\partial^{\nu}A^{\mu}(x),$
where $\partial^{\mu}=\frac{\partial}{\partial x_{\mu}}.$ Hence, the first
pair of the Maxwell equations, $\mathbf{\nabla\times B=}0,$ $\mathbf{\nabla
\times E+}\partial^{0}\mathbf{B}=0,$ with the electric and magnetic field
strengths $E_{i}=F^{i0}$ and $B_{i}=\epsilon_{ijk}F_{jk},$ remains standard.
The self-coupling constant $\gamma$ is presumably small enough. It has
dimension of inverse fourth power of mass.

The causality and unitarity principles applied to the local effective action
of any nonlinear electrodynamics result in some requirements  \cite{Shabus},
of which the first one is  $\gamma>0.$ Then, the other requirements (in the
present case, where the second field invariant $\mathfrak{G}=\mathbf{E\cdot
B}$ is not involved) reduce to $\frac{\partial L}{\partial\mathfrak{F}%
}\leqslant0,$ and $\frac{\partial L}{\partial\mathfrak{F}}+\frac{\partial
^{2}L}{\partial\mathfrak{F}^{2}}\leqslant0.$ These are satisfied up to an
infinite field strength, provided that $\mathfrak{F}<0.$ We shall be dealing
with this case of electric field taken alone in the present paper. This is a
certain advantage over the Born-Infeld model that becomes inconsistent at too
large electric field value.

It is known that the Born-Infeld model is a unique -- in the class of local
theories -- Lorentz- and gauge-invariant nonlinear theory, which is free of
birefringence \cite{Pleb} in propagation of small electromagnetic waves
against an external field background. The model (\ref{L}) does not share this
exceptional, but still not obligatory, property. Instead, it leads to
coincidence between transverse and longitudinal --with respect to the external
field -- dielectric constants as these are defined in \cite{Shabus}. (This is
not the birefringence yet, because the magnetic sector remains anisotropic,
unlike the electric one.)  

We are aiming at consideration of quantum electrodynamics (QED) as a theory,
whose intrinsic nonlinearity, originating from quantum corrections to the
linear classical Maxwell theory, is given, in the local approximation, by the
Euler-Heisenberg effective action functional. Correspondingly, the scale of
our coupling constant $\gamma$ will be emulated to QED, to be more precise, it
will be taken over from the light-by-light scattering amplitude of soft
long-wave photons.

\section{Field equations}

The least action principle applied to the functional $S\left[  A\right]  =\int
L(x)d^{4}x$ provides equations of motion $\frac{\delta S}{\delta A_{\nu}%
(x)}=0$, which are the second pair Maxwell equations, which are nonlinear:

\ \ \ \ \ \ \ \ \ \ \ \ \ \ \ \ \ \ \ \ \ \ \ \ \ \ \ \ \ \ \ \ \ \ \ \ \ \ \ \ \ \ \ \ \ \ \ \ \ \ \ \ \ \ \ \ \ \ \ \ \ \ \ \ \ \ \ \ \ \ \ \ \ \ \ \ \ \ \ \ \ \ \ \ \ \ \ \ \ \ \ \ \ \
\begin{equation}
\partial_{\mu}\left[  (1-\gamma\mathfrak{F}\left(  x\right)  )F^{\mu\nu
}\right]  =0. \label{Maxwell equations}%
\end{equation}
We are interested in purely electrostatic spherically symmetric solution \ to
this equation produced by a pointlike static charge $e$, placed in the origin.
Then this equation is reduced to%

\begin{equation}
\mathbf{\nabla}\left[  \left(  1+\frac{\gamma}{2}E^{2}\right)  \mathbf{E}%
\right]  =0 \label{Electrostatic equation1}%
\end{equation}
everywhere, except the point $\mathbf{x}=0.$ Bearing in mind that at large $r$
the standard Coulomb field of the point charge $e$%
\begin{equation}
\mathbf{E}\left(  \mathbf{x}\right)  =\mathbf{E}^{\text{lin}}\left(
\mathbf{x}\right)  =E^{\text{lin}}\left(  r\right)  \frac{\mathbf{x}}{r}%
=\frac{e}{4\pi r^{2}}\frac{\mathbf{x}}{r},
\end{equation}
where \ $r=|\mathbf{x|}$, should be implied as the boundary condition, we
rewrite (\ref{Electrostatic equation1}), up to a curl, as\ \
\begin{equation}
\left(  1+\frac{\gamma}{2}E^{2}(r)\right)  \mathbf{E}\left(  \mathbf{x}%
\right)  =\mathbf{E}^{\text{lin}}\left(  \mathbf{x}\right)  \mathbf{.}
\label{Electrostatic equation2}%
\end{equation}
In understanding that the coordinate $\mathbf{x}$ is the only vector in the
central-symmetric problem we may write $\mathbf{E}\left(  \mathbf{x}\right)
=E(r)\frac{\mathbf{x}}{r}.$ Then the mentioned curl must be discarded, because
it cannot be formed with $\mathbf{x}$ being the only vector, the first Maxwell
equation $\mathbf{\nabla\times E=}0$ is trivially satisfied.

\bigskip

\section{An exact solution to the quartic model}

\bigskip

Now equation (\ref{Electrostatic equation2}) becomes the cubic equation for
$E(r)$ (\textit{cf.} the procedure in \cite{CosGitSha2013})%

\begin{equation}
\left(  1+\frac{\gamma}{2}E^{2}(r)\right)  E\left(  r\right)  =\frac{e}{4\pi
r^{2}},\label{Cubic equation}%
\end{equation}
whose solution is given by the Cardan formula as%
\begin{equation}
E\left(  r\right)  =\sqrt[3]{\frac{E_{\text{lin}}\left(  r\right)
}{\mathfrak{\gamma}}+\sqrt{\left(  \frac{E_{\text{lin}}\left(  r\right)
}{\mathfrak{\gamma}}\right)  ^{2}+\left(  \frac{2}{3\mathfrak{\gamma}}\right)
^{3}}}-\sqrt[3]{\sqrt{\left(  \frac{E_{\text{lin}}\left(  r\right)
}{\mathfrak{\gamma}}\right)  ^{2}+\left(  \frac{2}{3\mathfrak{\gamma}}\right)
^{3}}-\frac{E_{\text{lin}}\left(  r\right)  }{\mathfrak{\gamma}},}\label{E}%
\end{equation}
which is its only real solution. For large distances, $r\rightarrow\infty,$
solution (\ref{E}) behaves itself as the standard Coulomb field $E\left(
r\right)  \sim E^{\text{lin}}\left(  r\right)  =\frac{e}{4\pi r^{2}},$ which
corresponds to neglect of the nonlinear quadratic term inside the bracket in
(\ref{Cubic equation}), because it is much less than unity in this limit. For
short distances, $r\rightarrow0,$ the asymtote of solution (\ref{E}) is%
\begin{equation}
E\left(  r\right)  \sim\left(  \frac{2E_{\text{lin}}\left(  r\right)
}{\mathfrak{\gamma}}\right)  ^{\frac{1}{3}}=\left(  \frac{e}{2\pi
\mathfrak{\gamma}}\right)  ^{\frac{1}{3}}\left(  \frac{1}{r}\right)
^{\frac{2}{3}},\label{Enear origin}%
\end{equation}
which might also be immediately obtained from (\ref{Cubic equation}) if we
neglected the unity in its left-hand side in favor of the quadratic term, much
larger than unity in this limit. The behavior of the electrostatic field
(\ref{Enear origin}), produced by the point charge $e$ via the nonlinear field
equations (\ref{Maxwell equations}), is essentially less singular in the
vicinity of the charge than the standard Coulomb field $E^{\text{lin}}\left(
r\right)  =\frac{e}{4\pi r^{2}}.$ We shall see below that this suppression of
the singularity is enough to provide convergence of the integrals giving the
energy of the field configuration (\ref{E}). Note that, contrary to the
customary situation \cite{Born} in  the Born-Infeld model, the singularity in
our case is not totally removed, but only suppressed to a sufficient extent. 

\bigskip

\subsection{Finiteness of the field energy of a point charge}

\ \ The Noether energy-momentum tensor for the Lagrange density (\ref{L}) is%
\begin{equation}
T^{\rho\nu}=(1-\gamma\mathfrak{F}\left(  x\right)  )F^{\mu\nu}\partial^{\rho
}A_{\mu}-\eta^{\rho\nu}L(x).
\end{equation}
By subtracting the full derivative $\partial_{\mu}\left[  \left(
1-\gamma\mathfrak{F}\left(  x\right)  F^{\mu\nu}\right)  A^{\rho}\right]
,$\ equal to $\left[  \left(  1-\gamma\mathfrak{F}\left(  x\right)  F^{\mu\nu
}\right)  \partial_{\mu}A^{\rho}\right]  $ due to the field equations
(\ref{Maxwell equations}), the gauge-invariant and symmetric under the
transposition $\rho\leftrightarrows\nu$ energy-momentum tensor\bigskip%
\[
\Theta^{\rho\nu}=(1-\gamma\mathfrak{F}\left(  x\right)  )F^{\mu\nu}F_{\mu
}^{\ \ \rho}-\eta^{\rho\nu}L(x)
\]
is obtained. When there is only spherically symmetric electric field, the
energy density is%
\begin{equation}
\Theta^{00}=(1+\frac{\gamma E^{2}}{2})E^{2}-\frac{E^{2}}{2}\left(
1+\frac{\gamma E^{2}}{4}\right)  =\frac{E^{2}}{2}+\frac{3\gamma E^{4}}{8}.
\label{Energy density}%
\end{equation}
By multiplying (\ref{Cubic equation}) by $E$ we obtain the relation
$\frac{\gamma}{2}E^{4}(r)=$ $E^{\text{lin}}\left(  r\right)  $ $E\left(
r\right)  -E^{2}\left(  r\right)  .$ Taking it into account the energy density
becomes%
\[
\Theta^{00}=\frac{E^{2}}{2}+\frac{3}{4}\left(  E^{\text{lin}}\left(  r\right)
E\left(  r\right)  -E^{2}\left(  r\right)  \right)  =-\frac{E^{2}\left(
r\right)  }{4}+\frac{3}{4}E^{\text{lin}}\left(  r\right)  E\left(  r\right)
.
\]
Therefore, in order to determine the full electrostatic energy $\int%
\Theta^{00}$d$^{3}x$ stored in solution (\ref{E}) we have to calculate two
integrals. The first one is\bigskip%
\[
\int E^{2}\left(  r\right)  \text{d}^{3}x=|e|^{\frac{3}{2}}\left(  \frac
{3}{2\gamma\left(  4\pi\right)  ^{2}}\right)  ^{\frac{1}{4}}\frac{3}{2}I_{1},
\]

\bigskip where\bigskip%
\begin{equation}
I_{1}=\int_{0}^{\infty}y^{\frac{2}{3}}\left(  \sqrt[3]{\sqrt{1+y^{4}}%
+1}-\sqrt[3]{\sqrt{1+y^{4}}-1}\right)  ^{2}\text{d}y=0.885.\nonumber
\end{equation}
The second one is%
\[
\int E^{\text{lin}}\left(  r\right)  E\left(  r\right)  \text{d}^{3}x=e%
{\displaystyle\int\limits_{0}^{\infty}}
E\left(  r\right)  \text{d}r=|e|^{\frac{3}{2}}\left(  \frac{3}{2\gamma\left(
4\pi\right)  ^{2}}\right)  ^{\frac{1}{4}}I_{2},
\]
where
\[
I_{2}=\int_{0}^{\infty}y^{-\frac{2}{3}}\left(  \sqrt[3]{\sqrt{1+y^{4}}%
+1}-\sqrt[3]{\sqrt{1+y^{4}}-1}\right)  \text{d}y=3.984.
\]
Finally the energy is%
\begin{equation}
\int\Theta^{00}\text{d}^{3}x=|e|^{\frac{3}{2}}\left(  \frac{3}{2\gamma\left(
4\pi\right)  ^{2}}\right)  ^{\frac{1}{4}}\frac{1}{4}\left(  3I_{2}-\frac{3}%
{2}I_{1}\right)  =2.65|e|^{\frac{3}{2}}\left(  \frac{3}{2\gamma\left(
4\pi\right)  ^{2}}\right)  ^{\frac{1}{4}}<\infty. \label{final energy}%
\end{equation}

\section{Polynomial model}

It is straightforward to extend the above statement about the finiteness of
the field energy of a point charge to any nonlinear electrodynamics, with the
effective Lagrange density $\mathfrak{L(\mathfrak{F})}$ \ -- in place of the
quartic function $\frac{\gamma}{2}\left(  \mathfrak{F(}x)\right)  ^{2}$ used
in (\ref{L}) -- being any function of $\mathfrak{F}$\ that grows as a finite
power of its argument, say $\mathfrak{F}^{n+1},$ $n\geqslant1,$ when
$\mathfrak{F\rightarrow\infty.}$ Neglecting again the unity in the equation to
appear in place of (\ref{Cubic equation}) we obtain in place of
(\ref{Enear origin}) that the electric field's singularity near the origin is%
\begin{equation}
E\left(  r\right)  \sim\left(  \frac{en!(-1)^{n+1}}{2^{n}4\pi\mathfrak{L}%
^{(n+1)}r^{2}}\right)  ^{\frac{1}{2n+1}}, \label{Enear origin2}%
\end{equation}
where $\mathfrak{L}^{(n+1)}$ is the $(n+1)$-st derivative of $\mathfrak{L}$
taken at $\ \mathfrak{F=}$ $0$. On the other hand, the leading-in-the origin
contribution to the field energy density calculated as Noether's $\Theta^{00}$
will now, instead of $E^{4}$ in (\ref{Energy density}), be proportional to
$E^{2n+2}.$ In spite of this higher power, the integral for the field energy
$\int\Theta^{00}$d$^{3}x$ with the substitution of (\ref{Enear origin2})
converges at the lower limit as $\int_{0}r^{-\frac{2}{2n+1}}$d$r,$ i.e., even
faster than that of (\ref{Energy density}). As for convergence at large
distances, it is ever provided by the standard Coulomb long-range behavior of
any nonlinear solution with the long-range boundary condition $E\left(
r\right)  \sim E^{\text{lin}}\left(  r\right)  ,$ when all nonlinearity in the
equation of motion should be disregarded..

The remark of the previous paragraph results in the claim that in QED, if one
truncates (like in \cite{CosGitSha2013}) the Taylor series expansion of its
nonlinearity at any given power of the field invariant $\mathfrak{F,}$ the
solution of the corresponding nonlinear Maxwell equations for electrostatic
field of a point charge is a finite-energy field configuration. It is meant
the effective action of QED defined as the generating functional of the
one-particle-irreducible vertex functions, or the Legandre transform of the
generating functional of the photon Green functions \cite{Weinberg}, is taken
in the local, or infrared, approximation\cite{Shabus, CosGitSha2013,
GitSha2012} . It may be thought of as the Euler-Heisenberg action calculated
with the accuracy of any number of loops.

\section{Towards field-mass of electron in QED as a nonlinear theory}

To estimate the result (\ref{final energy}) we may substitute the value of the
coupling constant $\gamma$ in (\ref{L}) taken equal to the coefficient by the
corresponding quartic term in the expansion of the one-loop Euler-Heisenberg
Lagrangian density $\mathfrak{L}^{\text{EH}}\mathfrak{,}$ i.e. (note that
$\alpha=\frac{e^{2}}{4\pi}=\frac{1}{137}$ in the Heaviside-Lorentz system used
here) \cite{BerLifPit}
\begin{equation}
\gamma=\frac{\partial^{2}\mathfrak{L}^{\text{EH}}}{\partial
\mathfrak{\mathfrak{F}}^{2}}=\frac{e^{4}}{45\pi^{2}m^{4}}, \label{gamma}%
\end{equation}
where $m$ is the electron mass, and $e$ is the electron charge in
understanding that the point charge previously denoted by the same letter now
also belongs to the electron. With this substitution the static field energy
(\ref{final energy}) of the electron considered as a purely electric
point-like monopole gives the result%

\begin{equation}
\int\Theta^{00}\text{d}^{3}x=2.09m, \label{mass}%
\end{equation}
about twice as big as the electron mass.

Are there prospects for making this value closer to $m$ in order to meet the
idea traced back to Abraham-Lorentz \cite{Abraham-Lorentz} and most advanced
in Born-Infeld electrodynamics \cite{Born} of an electron being a
particle-like finite-energy field configuration, the soliton, whose energy
would be of a completely field nature$?$ It is already a surprise that the
most trivial, quartic, term in the Lagrange density expansion might have given
the order-of-magnitude coincidence (\ref{mass}). Higher powers of nonlinearity
converge faster and faster as the power grows, and produce some corrected
values to replace (\ref{mass}). Although these terms depend on Feynman
diagrams with six, eight and more even-number prongs, the corrected values of
(\ref{mass}) will differ by more than powers of the fine structure constant:
note that Eq. (\ref{final energy}) with (\ref{gamma}) is of the order of
$\sqrt{|e|}$ , so we do not face a perturbative series. On the other hand, the
realistic electron, besides being an electric monopole, is also a pointlike
magnetic dipole, so the associated magnetic field energy should be expected to
contribute to the total field mass \cite{Blinder}. It can be shown that the
magnetic energy stored in the corresponding magnetic dipole field, if taken
independently of the electric field, is also converging in the model
(\ref{L}). A more challenging problem is to take the both mutually interacting
fields together.

\section*{Acknowledgements}

C. Costa acknowledges the support of CAPES. Gitman thanks CNPq and FAPESP for
permanent support. Besides, his work is done partially under the project
2.3684.2011 of Tomsk State University. A. Shabad acknowledges the support of
RFBR under the Project 11-02-00685-a. He also thanks USP for kind hospitality
extended to him during his stay in Sao Paulo, Brazil, where this work was started.


\begin{thebibliography}{9}                                                                                                %


\bibitem {Shabus}A.E. Shabad and V.V. Usov, \emph{Effective Lagrangian in
nonlinear electrodynamics and its properties of causality and unitarity,
}Phys. Rev. D \textbf{83, }105006 (2011).

\bibitem {Pleb}G. Boillat, \emph{Nonlinear electrodynamics: Lagrangians and
equations of motion,} J. Math. Phys. \textbf{11}, 941 (1969); J.
Pleba\'{n}ski,\emph{ Lecture notes on nonlinear electrodynamics}, (NORDITA,
Copenhagen, 1970).

\bibitem {CosGitSha2013}C. V. Costa, D.M. Gitman, and A.E.Shabad,\emph{
Nonlinear corrections in basic problems of electro- and magneto-statics in the
vacuum,} Phys. Rev. D \textbf{88}, 085026 (2013),\emph{\ }arXiv:1307.1802
[hep-th] (2013).)

\bibitem {Born}M. Born and L. Infeld, \emph{Electromagnetic mass,} Nature,
\textbf{132}, 970 (1933)).

\bibitem {Weinberg}{S. Weinberg, \emph{The Quantum Theory of Fields}
(University Press, Cambridge, 2001})

\bibitem {GitSha2012}D. M. Gitman and A. E. Shabad, \emph{Nonlinear (magnetic)
corrections to the field of a static charge in an external field, }Phys. Rev.
D \textbf{86}, 125028 (2012); arXiv:1209.6287; T.C. Adorno, D. M. Gitman, and
A. E. Shabad, \emph{Magnetic response to applied electrostatic field in
external magnetic field, }arXiv:1311.4081[hep-th].

\bibitem {BerLifPit}V. B. Berestetsky, E. M. Lifshits, and L. P. Pitayevsky,
\emph{Quantum Electrodynamics} (Nauka, Moscow, 1989; Pergamon Press, Oxford,
New York, 1982).

\bibitem {Abraham-Lorentz}Lorentz, H.A., \emph{Weiterbildung der Maxwell'schen
Theorie: Elektronentheorie.,} Encyklop\"{a}die d. Mathematischen
Wissenschaften, Band V2, Heft 1, Art. 14, 145-288 (1904); Abraham, M.,
\emph{Theorie der Elektrizit\"{a}t}, \emph{II}, Teubner, Leipzig (1905,1923);
Rohrlich, F., \emph{Classical charged particles}, Addison Wesley, Redwood
City, CA (1990); Spohn, H.,\emph{ Dynamics of charged particles and their
radiation field}, Cambridge University Press, Cambridge (2004).
\end{thebibliography}
\end{document}